\documentstyle[aps,twocolumn,psfig]{revtex}
\draft
\preprint{Submitted to {\it Physical Review Letters} \hspace{7cm} LA-UR-97-3040}
\newcommand{\etal}     {{\it et al}}
\newcommand{\lc}       {lanthanum cuprate}

\newcommand{\lilco}    {${\rm La_2 Cu}_{1-x} {\rm Li}_{x} {\rm O_4}$}
\newcommand{\lco}      {$\rm La_2 Cu O_4$}
\newcommand{\lsco}     {${\rm La}_{2-y} {\rm Sr}_y {\rm Cu O_4}$}
\newcommand{\cuot}     {CuO$_2$}

\newcommand{\tn}       {$T_N$}
\newcommand{\ms}       {$M_s$}
\newcommand{\mso}      {$M_s^0$}
\newcommand{\la}       {$^{139}$La}
\newcommand{\rb}[1]    {\raisebox{1.5ex}[0pt]{#1}}

\title{Suppression of Antiferromagnetic Order by 
 Light Hole Doping in La$_2$Cu$_{1-x}$Li$_x$O$_4$: \\ A $^{139}$La NQR Study}
\author{B. J. Suh,$^1$ P. C. Hammel,$^1$ Y. Yoshinari,$^1$ J. D. Thompson,$^1$ 
J. L. Sarrao,$^1$ 
and Z. Fisk$^2$}
\address{$^1$Condensed Matter and Thermal Physics, 
Los Alamos National Laboratory, Los Alamos, NM 87545 \\
$^2$National High Magnetic Field Laboratory, 
Florida State University, Tallahassee, FL 32306\vspace{-2mm}}
\author{\small(Received:\qquad  October 22, 1997)}

\address{
\parbox{14cm}{\bigskip\rm\small
\la\ nuclear quadrupole resonance measurements in
lightly doped \lilco\ have been performed to reveal the 
dependence of the magnetic properties of the antiferromagnetic \cuot\ 
planes on the character of the doped holes and their interactions with the dopant.  
A detailed study shows that the magnetic properties are remarkably insensitive to the 
character of the dopant impurity.  
This indicates that the added holes form previously unrecognized collective 
structures. 
}}

\begin{document}
\maketitle

\thispagestyle{myheadings}
\markright{{\em LA-UR-97-3040} \hspace{40mm}{\small {\em Physical Review Letters} (accepted)} \hfill \hspace{1mm}}

Full understanding of the character of holes added to cuprate planes 
and their interactions with the two-dimensional lattice of Cu spins 
remains a crucial and unsolved problem in the high temperature superconductors. 
While the detailed mechanism is poorly understood, 
the rapid suppression of the antiferromagnetic (AF) ordering temperature $T_N$
by doping is clearly related to the disruptive effects of mobile holes:
$\lesssim 3\%$ added holes whether from Sr substitution, 
addition of interstitial oxygen or 
in-plane substitution of Li for Cu\cite{Sarrao} suppresses \tn\ to zero, 
yet $\sim 30\%$ isovalent substitution of Zn or Mg for Cu is required\cite{Zn1} 
to produce the same effect.  
Li and Sr-doped holes have very different mobilities. 
For $x \text{ or } y \approx 0.025$, 
the room temperature resistivity $\rho$ of 
\lilco\ exceeds that of \lsco\ by over an 
order of magnitude\cite{Sarrao,kastner,takagi}; 
more strikingly,  for Sr-doping $d\rho/dT>0$ for $T\gtrsim 100\,$K,
in contrast to the negative slope found in Li-doped material for 
all $x$ and $T$. 

It is well recognized that the 2D cuprates are inclined 
toward microscopic charge inhomogeneity\cite{Emery,zaanen,latteff}. 
Evidence for such an effect in lightly doped \lsco\ was obtained from a 
scaling analysis of the doping $y$ and temperature $T$ dependence of the static 
susceptibility $\chi (y,T)$\cite{cho1} which indicated that the 
magnetic correlation length is limited to the dimensions of AF domains 
(finite-size scaling) formed by microsegregation of doped holes  
into hole-rich domain walls surrounding hole-free, AF domains.  
Interpretations involving charge-stripes have also been proposed\cite{Borsa}. 
Castro Neto and Hone\cite{neto} have examined the influence 
of doping on the long wavelength properties of a 2D antiferromagnet 
in a model in which charged stripes 
cause the exchange coupling $J$ to become anisotropic;  
this model reproduces the relationship between \mso\ 
(\ms\ is the sublattice magnetization; \mso\ is that obtained by extrapolation 
of data for $T>30\,$K to $T=0$) 
and \tn\ as the two are suppressed by Sr-doping
in \lsco\cite{Chou}.
However, using a similar model, van Duin and Zaanen\cite{vdz} 
find that \tn\ is suppressed much more rapidly than \mso\ with increasing
anisotropy (doping).

We have used \la\ nuclear quadrupole resonance (NQR) measurements to  
microscopically examine the effects of doped holes on the AF spin correlations 
in \lilco\ ($ 0.019 < x < 0.025$).  
We find that the magnetic behavior 
of \lc\ is remarkably insensitive to the detailed nature of the dopant, 
in spite of the differing charge transport associated with the two dopants. 
In addition to the similarly strong suppression of \tn\ by doping, 
we find the identical correspondence between the suppression 
of \mso\ and \tn\ by doping which has been observed in \lsco.  
Further we show, for the first time, that in the vicinity of \tn\  
the dynamical susceptibility, 
as reflected in the nuclear spin relaxation rate $2W$, 
follows a scaling law consistent with the finite-size scaling 
demonstrated in the static susceptibility by Cho \etal. \cite{cho1}.
Finally, at low temperature, we find that two peculiar features are 
very similar in the two systems.
These are the very strong peak in $2W$ 
(at a temperature $T_f=10$--16 K depending on $x$) 
that indicates freezing of spin-degrees of freedom,  
which is accompanied at slightly higher temperatures ($\simeq 30\,$K) 
by the abrupt recovery of $M_s(T)$, almost to $x=0$ values.  

Thus, while very small concentrations of added holes induce a range of 
characteristic magnetic properties which are entirely insensitive to 
the nature of the dopant, 
the transport properties are very sensitive to the dopant.
Unable to understand these contrasting behaviors as arising from 
properties of individual holes, 
we conclude holes form collective structures. 
We will argue that holes form charged, anti-phase domain walls\cite{tranq} 
which surround mobile domains in which the phase of the AF order is reversed.
Such mobile domains will suppress the time-averaged static moment thus 
suppressing AF order and \ms.  
These domain structures will have contrasting interactions with in-plane 
{\it vs.\ }out-of-plane dopants 
({\it e.g.,\/} stronger scattering by in-plane impurities) 
which explain the different transport behaviors, 
while the universal magnetic properties can be understood 
as long as the domains are sufficiently mobile that they move across a given site 
rapidly compared to a measurement time.

Three powder samples of \lilco\ (labeled A1, B1 and B2)   
were prepared from starting material containing concentrations $x_{\rm nom}$ of Li 
as described elsewhere\cite{Sarrao}.  
Table~\ref{table1} shows various measured properties of these samples;
definitions of the parameters shown and the means by which they were 
determined will be discussed in what follows.
All the samples were annealed in flowing
nitrogen gas after sintering to minimize excess oxygen,  
whose motion can contribute to $^{139}$La spin-lattice relaxation at high
$T$ \cite{Chou,Silvia}.
Annealing in nitrogen eliminated this contribution in our samples.
The smaller value of $x$ relative to $x_{\rm nom}$ for sample A1 
was deduced from its higher value of \tn\ relative to sample B1. 
$^{139}$La $\left(I = {7\over 2}\right)$ NQR and relaxation rate measurements in
\lilco\ were performed for 4~K$\,<T<\,$300~K on both the 
$2\nu_Q \left (\pm {5\over 2} \! \leftrightarrow \! \pm {3\over 2} \right)$ and 
$3\nu_Q \left (\pm {7\over 2} \! \leftrightarrow \! \pm {5\over 2} \right)$ 
transitions.  
The spectra were obtained by
plotting the integrated intensity of the spin-echo signal 
as a function of spectrometer frequency.  The nuclear spin-lattice
relaxation rate was measured by monitoring the recovery of the 
magnetization after saturation with a single $\pi \over 2$ pulse.

\begin{figure}
\parbox{1mm}{\rule{0mm}{5mm}}
\parbox{3in}{
\psfig{file=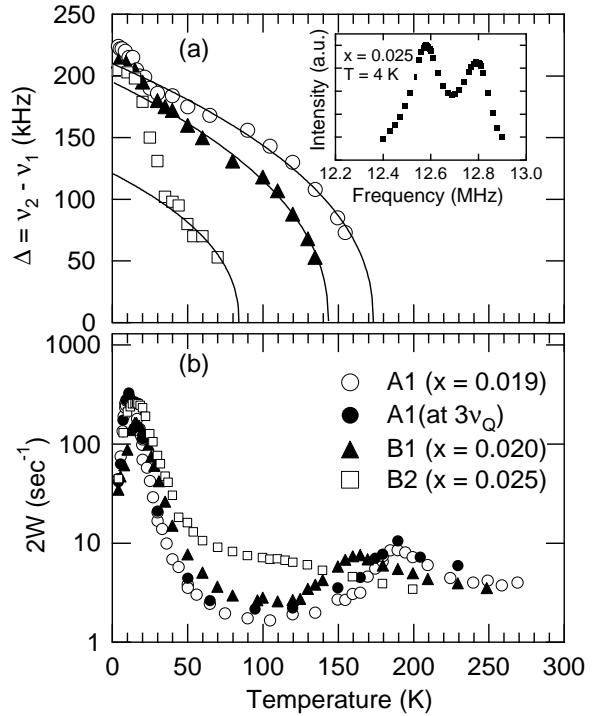,width=3.4in}}
\vspace{2mm}
\caption{$^{139}$La NQR in La$_2$Cu$_{1-x}$Li$_x$O$_4$:
(a)~$\Delta \equiv \nu _1 - \nu _2$ {\it vs.\/}$\,\,T$.  Solid
curves are fits to the 
behavior $\Delta(T)=\Delta
_0(1-T/T_N)^{\beta}$, $\beta = 0.44 \pm 0.01$.  
The inset shows a representative NQR spectrum at 4 K. 
(b)~$2W$ {\it vs.\ }$T$.}
\label{fig:dnu}
\end{figure}

The $^{139}$La NQR results for 
the three samples are summarized in Fig.~\ref{fig:dnu}.  
Below \tn, the ordered Cu moment generates an internal magnetic 
field {\bf H}
at the $^{139}$La site which splits the NQR line into a doublet with
frequencies $\nu _1$ and $\nu _2$.  
The magnitude of the splitting, $\Delta \equiv \nu _1 - \nu _2$, 
is a direct measure of the component
$H_z=H\cos{\theta}$ of the internal field {\bf H} at the La site along
the principal ($\hat z$) axis of the electric field gradient (EFG) 
tensor\cite{Borsa} ($\theta$ is the angle between {\bf H} and $\hat z$).
The two frequencies for the $2\nu_Q$ transition are 
given by $\nu _{1,2} = 2\nu _Q \pm (\gamma _n/2 \pi)H_z$ 
where the nuclear gyromagnetic ratio $\gamma _n/ 2\pi=601.44$~Hz/G.  
As shown in Fig.~\ref{fig:dnu}(a), for $T>30\,$K, 
$\Delta$ is suppressed by doping, and then displays an abrupt increase below 30~K.  
Similar effects have been observed in recent $\mu$SR and neutron scattering 
measurements in \lilco\cite{lh}.
The solid curves in Fig.~\ref{fig:dnu}(a) are fits of the temperature dependence 
$\Delta(T)=\Delta_0(1-T/T_N)^{\beta}$ 
to the data for $T> 30$~K; we find $\beta = 0.44 \pm 0.01$.
The magnitude and temperature dependence of $\nu_Q$ for a 
given doping is almost identical to that found in 
Ref.~\onlinecite{Chou};  
the linewidths are also quite comparable:
110, 120, and 140 kHz at 250 K for our samples A1, B1 and B2, 
respectively. 

Fig.~\ref{fig:dnu}(b) shows the $T$-dependence of $2W$, which displays a
strong peak at $T= T_f=10\,$--16 K and a weak 
peak in the vicinity of $T_N$.
$2W$
was obtained by fitting the recovery data to the theoretical expression 
for magnetic relaxation and for single pulse saturation\cite{Mac}.
For $T > T_f$, this expression fits the data well, 
indicating a 
single rate arising from magnetic fluctuations. 
Although the rate becomes distributed below $T_f$
(consistent with freezing in an inhomogeneous distribution of internal 
\vspace{-3mm}
\begin{table}
\caption{Properties of the three samples: A1, B1 and B2.}
\label{table1}
\begin{tabular}{*{10}{c}}
&                   &         & $T_N$ & $E_a/k_B$ & $\Delta _0$ & $T_f$ & $T_N^*$ &          & $Cx^2$          \\ 
&\rb{$x_{\rm nom}$} & \rb{$x$}& (K)   &  (K)      & (kHz)       &  (K)  &    (K)  & \rb{$C$} & $\times 10^{5}$ \\ \hline
A1  & 0.020   & 0.019 & 180$\pm 5 $ & 120  & 212 & 11  & 187  & 0.042       & $1.52 $ \\ 
B1  & 0.020   & 0.020 & 145$\pm 5 $ & 123  & 193 & 15  & 163  & 0.038       & $1.52 $ \\
B2  & 0.025   & 0.025 &  85$\pm 10$ & 117  & 120 & 16  & 114  & 0.020       & $1.25 $
\end{tabular}
\end{table}
\noindent 
fields), 
the same fitting procedure was applied to data for the first decade of recovery. 
While this increases the uncertainty in $2W$, 
we find that varying the fitting procedure has essentially no 
effect on the position of the peak at $T_f$.

As seen in Fig.~\ref{fig:dnu}(b), 
$2W(T)$ is the same at the $2\nu _Q$ and $3\nu _Q$ transitions in sample A1, 
establishing that $2W$ is due to a magnetic rather than structural mechanism.
In addition, we note that $2W$ decreases with increasing $T$ 
for $T >T_N$ in contrast with the
results in \lsco\ which show an enhancement of $2W$ at high $T$ due
to the motion of excess oxygen \cite{Silvia}.  
Thus, annealing to remove excess oxygen enables us
to obtain, for the first time, intrinsic data for spin dynamics above $T_N$ 
that are isolated from any significant influence by mobile oxygen.

Our discussion will focus on two aspects of the data: (i) the anomalous
behavior of both the static and the 
dynamical magnetic properties at low $T$, 
and (ii) the broad and weak peak in $2W$ around $T_N$, 
which contrasts with the sharp peak at $T_N$ observed in, {\it e.g.,\/} undoped 
$\rm Sr_2 Cu O_2 Cl_2$\cite{Suh}.
These results are, overall, very similar to those found 
in \lsco\cite{Borsa,Chou}.  

\begin{figure}
\parbox{7mm}{\rule{0mm}{5mm}}
\parbox{3in}{
\psfig{file=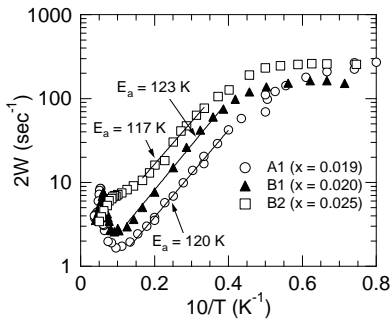,width=2.1in}}
\vspace{3mm}\caption{$2W$ {\it vs.\/} $10/T$.  Solid
lines are fits to the activated behavior $2W \propto \exp (E_a/k_BT)$.}
\label{fig:activated}
\end{figure}

The strong low-$T$ peak in $2W$ clearly indicates freezing 
of spin degrees of freedom. 
Analyzing the data in terms 
of activated behavior, 
[$2W(T) \propto \exp(E_a/k_BT)$] as shown 
in Fig.~\ref{fig:activated},  
gives values of $E_a$ (see Table \ref{table1}) similar 
to those 
in \lsco; where $E_a/k_BT_f=8.9$, and $T_f \simeq 11$
\vspace{-6mm}
\noindent
--16~K
$\Rightarrow E_a \simeq 100$--145~K\cite{Chou}.  
Thus, the spin freezing is independent of the dopant,  
and, consequently, spin freezing is not solely determined by 
binding of holes to the dopant. 
We note that $T_f(x)$ (Table \ref{table1}) does not satisfy the
empirical relation $T_f=(815~{\rm K})x$ obtained in Ref.~\cite{Chou}, 
but the relationship between $T_f$ and $T_N$ is the same in both systems. 
Such disagreement may arise from a discrepancy in actual
hole-doping level.

\begin{figure}
\parbox{6mm}{\rule{0mm}{5mm}}
\parbox{3in}{
\psfig{file=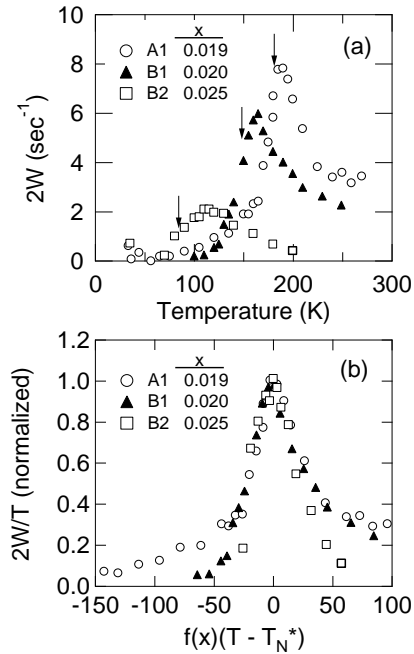,width=2.2in}}
\caption{(a)~$T-$dependence of $2W$: $2W$ data are corrected for the
contribution from the spin-freezing effects represented by the solid lines in
Fig~2.  The arrows indicate \tn . 
(b)~Plots of normalized $2W/T$ as a
function of $f(x)(T-T_N^*)$ with $f(x)=(0.02/x)^2$ (see text for
details).}
\label{fig:scaling}
\end{figure}

We turn now to the behavior of $2W$ in the vicinity of \tn.  
We find a strong doping dependence of the width of the $2W$ peak as 
found in the static susceptibility by Cho \etal.\cite{cho1},
and our results support the finite-size scaling proposed there.
To isolate the behavior near \tn\ from contamination by the tail of the 
spin-freezing peak at $T_f$, 
we first correct the $2W$ data by subtracting the fitting results in 
Fig.~\ref{fig:activated} (solid lines). 
Results are shown in Fig.~\ref{fig:scaling}(a).
The observed behavior cannot be understood in 
terms of conventional critical behavior\cite{Suh}.  
There, the $T$-dependent growth of $\xi$ is set by the exchange coupling 
constant $J$, 
and this determines the width of the peak on the high-$T$ side.
The strong $x$-dependence of the width then leads to the 
implausible conclusion that $J$ is strongly $x$-dependent.     
Nonetheless, it is clear that the behavior of $2W$
around $T_N$ should reflect cooperative behavior
of correlated spins near their ordering temperature.  

The peak in the static susceptibility observed at \tn\ is due 
to the small canting of the ordered moment which arises through 
the Dzyaloshinski-Moriya interaction.
In undoped \lco\ the shape of this peak is well described by theory\cite{thio}, 
however with doping the peak broadens and this description is no longer adequate.
Cho \etal.\ argued that rather than increasing exponentially 
with decreasing temperature, 
$\xi (T)$ in lightly doped \lsco\ is limited by confinement to AF 
domains of size $L$ defined by hole-rich domain walls\cite{cho1}. 
They were successful in describing the doping dependent shape 
of the peak in $\chi(y,T)$ (for $T>T_N$) for a range of $y$ 
through the scaling relation
$\chi(y,T) = \chi\{f(y)[T-T_N(y)]\}$,  
and they identify $f(y) \simeq 0.02/y^2$ with $L^2$.  
If this peak shape is indeed due to $\xi (T)$, 
consistency of the scaling of $2W/T$ and the static 
susceptibility is expected since both the static and dynamic 
susceptibilities ($\chi '$ and $\chi ''$, 
respectively) 
are determined 
by the same $T$-dependent $\xi$. 

We 
find that the peak in $2W(T)$ occurs at $T=T_N^*$ 
different from the peak in $\chi (T)$
[marked by arrows in Fig.~\ref{fig:scaling}(a); see also Table \ref{table1}].  
Given the incomplete understanding of $\chi (T,x)$ for finite $x$, 
the peak in $2W$ is likely the better indicator of the temperature at which 
local AF ordering occurs.  
Although not understood, we believe this phenomenon is a general 
feature of lightly doped \lc, and is more prominent in our data because 
the obscuring effects of mobile excess oxygen at temperatures 
comparable to \tn\ have been eliminated here.   
We find that $2W(T,x)$ 
exhibits the same scaling behavior
with the same scaling function $f(x)=(0.02/x)^2$. 
The strength of the peak in $2W$ is strongly $x$-dependent, 
and we apply a normalization factor $C(x)$:
\begin{equation}
{\textstyle {2W \over T}} \{x,T\}  = C(x) {\textstyle {2W \over T}} \{ f(x)[T-T_N^*(x)] \} .
\label{scaling}
\end{equation}
The data scaled in this way are shown in Fig.~\ref{fig:scaling}(b), 
and we find $C(x)\simeq (.004/x)^2$ as shown in Table \ref{table1}.  
Given the discrepancy between $T_N$ and $T_N^*$, 
these data cannot be 
taken as demonstrating scaling of $2W$; 
however, our finding that the same scaling function $f(x)$, 
when combined with a normalization factor also proportional to $1/x^2$, 
leads to consistent scaling behavior provides important corroboration 
for the finite-size scaling hypothesis of Cho \etal.\cite{cho1}. 

The reduction of $2W / T$ associated with 
limitation of the size of correlated domains is not unexpected.
The spin-lattice relaxation rate is given by
$2W/T \sim \sum_q A^2(q)[\chi''(q,\omega) / \omega]$,
where $A(q)$ is the (sample independent) hyperfine coupling constant. 
If the spectrum of 
\vspace{-3mm}
\noindent
spin fluctuations is described by a Lorentzian distribution with a characteristic frequency $\Gamma$, 
then $\chi''(q,\omega) / \omega \sim \chi'(q) / \Gamma $; 
this highlights the additional sensitivity of $2W$ to the characteristic fluctuation 
rate $\Gamma$ compared to the real part of the susceptibility. 
As the number of spins in a domain 
(argued in the finite-size hypothesis to go as $1/x^2$\cite{cho1}) 
increases, $\Gamma$ will decrease;  
and, thus, the relaxation rate will increase with decreasing doping. 
Qualitatively, this is observed:  
$2W/T \sim C(x) \sim  (1/x)^2 $.

The detailed similarity of the magnetic properties of 
Sr and Li-doped \lc\ cannot be understood in terms of individual holes
interacting with impurities; 
this suggests that holes form collective structures.
A full explanation of these magnetic properties 
cannot be given at this point, 
but to be concrete we propose a picture that can successfully account for the  
distinctive and complex behavior of $\Delta$ and $2W$ below \tn.   
Phase segregation of holes to boundaries of hole free regions \cite{cho1} or 
to hole rich ``rivers''\cite{Borsa} has been proposed. 
Based on the results of Tranquada \etal.\cite{tranq} at higher doping in 
$\rm La_{1.48} Nd_{0.4} Sr_{0.12} Cu O_4$, we suggest a contrasting picture 
in which the stripes are densely populated with holes  
(e.g., 1/2 hole per stripe Cu site) and constitute anti-phase domain walls.  
We further suggest that these stripes close to form mobile loops that 
surround small domains (which we call anti-phase 
domains) in the antiferromagnet. 
Such loops have been theoretically argued to be the stable configuration 
at low doping\cite{white:loops}.
Because these encircling stripes are anti-phase domain walls, 
the phase of the AF correlations in 
an antiphase domain is reversed with respect 
to the background. 
Motion of these domains (above 30 K)
over a particular site thus reverses the spin orientation 
and reduces the time-averaged static moment at that site. 

To estimate the size and density of domains required to explain experimental 
data, we consider a simple model\cite{hammel} in which there is, 
on average, one domain for every $N$ Cu sites (separated by lattice constant $a$).   
These $N$ sites occupy an area $L^2 =N a^2$, 
and, of these, $N_-$ sites comprise the anti-phase domain whose width is 
$l = a \sqrt{N_-}$.  
The fractional reduction of the time-averaged static moment at a given site 
will then be $(N_+ - N_-) / N$ where $N = N_+ + N_-$.
We set this equal to $R(x) \equiv M_s^0(x)/M_s^0(x=0)$. 
Assuming 1/2 hole per Cu in the stripe, we have $4l \sim 2xNa$ 
(we assume that all donated holes are incorporated into loops), 
and $l/a \sim (1-R)/x$. 
Using measured values of $R(y)$ for \lsco\cite{Borsa}, 
we find $l/a$ increases with doping to $\sim$ 20  
at $y=0.018$ where $R$ has decreased to $\sim 0.6$.
Recovery of the sublattice magnetization below 30 K occurs 
when the domains either become pinned to the lattice or 
evaporate as the constituent holes become pinned to donor impurities.
Mobile anti-phase domains will introduce disorder and reduce the total 
inter-plane coupling of AF ordered regions thus reducing \tn \cite{hammel}.

In conclusion, we have presented a complete set of $^{139}$La NQR data in
\lc\ doped by in-plane substitution of Li for Cu.  
The most striking result is the remarkable insensitivity of the magnetic 
properties of the AF \cuot\ planes to the nature and location of the dopant; 
this in spite of marked differences in transport properties.
This suggests the magnetic properties reflect collective hole 
phenomena.
The behavior of $2W$ around $T_N$ cannot 
be understood in the context of conventional critical behavior; 
we have demonstrated consistency with the finite-size 
scaling approach proposed earlier\cite{cho1}.  
We propose that all of the data can be understood by means of a model 
in which doped holes segregate to domain walls that enclose 
mobile, anti-phase bubbles which reduce the time-averaged ordered moment thus
suppressing \mso\ and \tn. 

We gratefully acknowledge stimulating conversations with F. Borsa, D. C. 
Johnston, A. H. Castro Neto and particularly J. Zaanen, 
who proposed the idea behind the model presented here. 
Work at Los Alamos performed under the auspices of the US Department of Energy.  
The NHMFL is supported by the NSF and the State of Florida through 
cooperative agreement DMR 95-27035.

\end{document}